\documentclass[aps,prl,twocolumn,superscriptaddress,showpacs]{revtex4}
\usepackage{amsfonts}
\usepackage{amssymb}
\usepackage{amsmath}
\usepackage{graphicx}
\usepackage{dcolumn}
\usepackage{bm}
\usepackage{flafter}

\begin{document}

\title{Quantum Oscillations of the Metallic Triangular-lattice Antiferromagnet PdCrO$_2$}

\author{Jong Mok Ok}
\affiliation{Department of Physics, Pohang University of Science and Technology, Pohang 790-784, Korea}
\author{Y. J. Jo}
\affiliation{Department of Physics, Kyungpook National University, Daegu 702-701, Korea}
\author{K. Kim}
\affiliation{Department of Physics, Pohang University of Science and Technology, Pohang 790-784, Korea}
\author{T. Shishidou}
\email{shishido@hiroshima-u.ac.jp}
\affiliation{ADSM, Hiroshima University, Higashihiroshima, Hiroshima 739-8530, Japan}
\author{E. S. Choi}
\affiliation{National High Magnetic Field Laboratory, Florida State University, Tallahassee, Florida 32310, USA}
\author{Han-Jin Noh}
\affiliation{Department of Physics, Chonnam National University, Gwangju 500-757, Korea}
\author{T. Oguchi}
\affiliation{ISIR, Osaka University, Ibaraki, Osaka 567-0047, Japan}
\author{B. I. Min}
\affiliation{Department of Physics, Pohang University of Science and Technology, Pohang 790-784, Korea}
\author{Jun Sung Kim}
\email{js.kim@postech.ac.kr}
\affiliation{Department of Physics, Pohang University of Science and Technology, Pohang 790-784, Korea}

\date{\today}

\begin{abstract}
We report the de Haas-van Alphen (dHvA) oscillations and first-principle calculations for triangular antiferromagnet PdCrO$_2$ showing unconventional anomalous Hall effect (AHE). The dHvA oscillations in PdCrO$_2$ reveal presence of several 2 dimensional Fermi surfaces of smaller size than found in nonmagnetic PdCoO$_2$. This evidences Fermi surface reconstruction due to the non-collinear 120$^{\rm o}$ antiferromagnetic ordering of the localized Cr, consistent with the first principle calculations. The temperature dependence of dHvA oscillations shows no signature of additional modification of Cr spin structure below $T_N$. Considering that the 120$^{\rm o}$ helical ordering of Cr spins has a zero scalar spin chirality, our results suggest that PdCrO$_2$ is a rare example of the metallic triangular antiferromagnets whose unconventional AHE can not be understood in terms of the spin chirality mechanism.
\end{abstract}


\maketitle

The two dimensional (2D) triangular-lattice antiferromagnet (TAFM) is one of the simplest frustrated magnets, but showing complex magnetic phases due to geometrical frustration. While most of TAFM systems are insulating with localized spins, there are a few metallic systems such as PdCrO$_2$~\cite{PdCrO2:maeno:AHE,PdCrO2:maeno:res,PdCrO2:Mekata:NPD,PdCrO2:maeno:NPD,PdCrO2:singh:cal}, AgNiO$_2$~\cite{AgNiO2:mazin:cal,AgNiO2:radaelli:npd,AgNiO2:coldea:dHvA}, Ag$_2$$M$O$_2$ ($M$ = Cr, Mn, Ni)~\cite{Ag2CrO2:yoshida:syn,Ag2MnO2:hiroi:syn,Ag2NiO2:hiroi:syn} and Fe$_{1.3}$Sb~\cite{Fe1.3Sb:tokura:AHE}. When itinerant electrons are coupled with intriguing magnetic orders in various frustrated magnets, unconventional transport behaviors are observed, $e.g.$ unconventional anomalous Hall effect (AHE) due to non-zero Berry phase associated with spin chirality~\cite{AHE:review,AHE:review2}. How complex magnetic order influences the nature of itinerant electrons or vice versa is, therefore, one of the key questions for understanding exotic properties of metallic frustrated magnets.

PdCrO$_2$ is of particular interest because it is a rare example of the TAFM's showing unconventional AHE~\cite{PdCrO2:maeno:AHE}. PdCrO$_2$ consists of stacked layers of Pd and Cr triangular lattices in a delafossite structure. The Pd layers with mostly Pd 4$d^9$ states are responsible for highly metallic conduction~\cite{PdCrO2:singh:cal} as found in the iso-structural non-magnetic compound PdCoO$_2$~\cite{PdCoO2:hjnoh:ARPES,PdCoO2:hjnoh:XAS,PdCoO2:min:band}. In the CrO$_2$ layer, three electrons of Cr$^{3+}$ ions in an octahedral environment fully fill the $t_{2g}^3$ states. In the absence of orbital ordering or structural distortion, the localized S = 3/2 spins of Cr$^{3+}$ ions are expected to be antiferromagnetically ordered in the 120$^{\rm o}$ helical spin structure with $\sqrt{3}\times\sqrt{3}$ periodicity. In fact, the 120$^{\rm o}$ helical ordering at $T_N$ $=$ 37.5 K has been suggested by neutron powder diffraction~\cite{PdCrO2:Mekata:NPD,PdCrO2:maeno:NPD}. For the ideal 120$^{\rm o}$ helical magnetic structure, however, the scalar spin chirality, defined as $S_i \cdot (S_j \times S_k)$, is canceled out, thus inconsistent with the observed unconventional AHE. In order to resolve the discrepancy, additional magnetic transition near $T^*$ $\sim$ 20 K has been proposed, which requires further investigations. In this respect, it is essential to clarify how the itinerant electrons showing unconventional AHE are coupled to the AFM ordering of the localized spins in the neighboring CrO$_2$ layer.

In this Letter, we present a study of de Haas-van Alphen (dHvA) effects and first principle calculations for PdCrO$_2$, evidencing significant coupling of itinerant electrons with the localized Cr spins. Several Fermi surfaces (FS) of PdCrO$_2$ are identified from the dHvA oscillations, which are much smaller than found in the nonmagnetic and iso-structural PdCoO$_2$~\cite{PdCoO2:mackenzie:dHvA}. This provides clear experimental evidence of FS reconstruction due to the 120$^{\rm o}$ AFM ordering with three magnetic sublattices, in good agreement with first principle calculations. Therefore, unlike the previous conjecture~\cite{PdCrO2:maeno:AHE}, the magnetic structure of PdCrO$_2$ is close to the ideal 120$^{\rm o}$ structure below $T_N$. Considering the zero scalar spin chirality of the ideal 120$^{\rm o}$ structure, our findings suggest that PdCrO$_2$ is a rare example of metallic triangular antiferromagnets whose unconventional AHE cannot be understood in terms of the spin chirality mechanism.

\begin{figure}
\includegraphics*[width=8.0cm, bb=10 80 565 805]{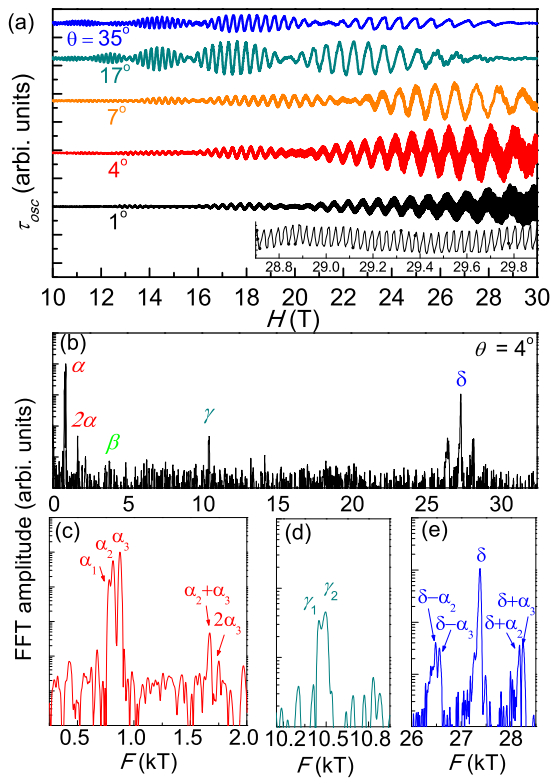}
\caption{\label{fig1}(color online) (a) The oscillatory part of the torque signal as a function of applied magnetic fields at 1.4 K for different angle $\theta$ with respect to the $c$ axis. The fast oscillation at high magnetic fields for $\theta$ = 1$^{\rm o}$ is shown in the inset. (b) The typical FFT spectra of the dHvA oscillations in the field range from 10 T to 30 T, for example, at $\theta$ = 4$^{\rm o}$. An expanded view of the FFT spectra for the different frequencies, labeled as (c) $\alpha$, (d) $\gamma$ and (e)$\delta$. Each FFT peak consists of more than two closely-spaced peaks (see the text).
}
\end{figure}

Single crystals of PdCrO$_2$ were grown by a flux method as describe in Ref.[\onlinecite{PdCrO2:maeno:syn,supp}]. For the torque measurements, a small single crystal, typically 50$\times$50$\times$10 $\mu$m$^3$ was mounted onto a miniature Seiko piezo-resistive cantilever. In total, five crystals were investigated, two in 14 T PPMS at POSTECH, two in 18 T superconducting magnet and one in 33 T Bitter magnet at National High Magnetic Field Lab. (NHMFL), Tallahassee, USA. All the crystals show consistent behaviors, and here we present the dHvA results obtained using 33 T magnet at NHMFL. For the first principle calculations, we employed the full potential linearized augmented plane wave (FLAPW) method using the HiLAPW code\cite{cal0} and also using the Wien2k-NCM code~\cite{cal}. For the exchange correlation potential, we have used the generalized gradient approximation (GGA)~\cite{supp}.

Figure 1(a) shows the oscillatory part of the torque signal up to 30 T for PdCrO$_2$ with several orientations as the magnetic field is rotated from $H$ $\parallel$ [001] ($\theta$ = 0$^{\rm o}$) towards $H$ $\parallel$ [100] ($\theta$ = 90$^{\rm o}$) at $T$ = 1.4 K. The fast Fourier transform (FFT) for $\theta$ = 5$^{\rm o}$ close to $H$ $\parallel$ $c$ is shown in Fig. 1(b). At all angles, the spectrum is dominated by several peaks at $F$ $\sim$ 0.8 kT, $\sim$ 3.3 kT, $\sim$ 10.5 kT and $\sim$ 27.5 kT, which we denote as $\alpha$, $\beta$, $\gamma$, and $\delta$, respectively.  The $\alpha$($\gamma$) branches consist of 3(2) closely-spaced peaks as labeled as $\alpha_i$($\gamma_i$) [Fig. 1(c) and (d)]. Also we observed 2nd harmonics of the $\alpha$ branches  at $F$ $\sim$ 1.6 kT [Fig. 1(c)]. The $\beta$ branches near $\sim$ 3.3 kT exhibit a complex structure consisting of several small peaks. Also for the $\delta$ branches, we observed mixing with $\alpha_i$ branches, resulting the $\delta$$\pm$$\alpha_i$ peaks.

\begin{figure}
\includegraphics*[width=8.0cm, bb=19 500 320 745]{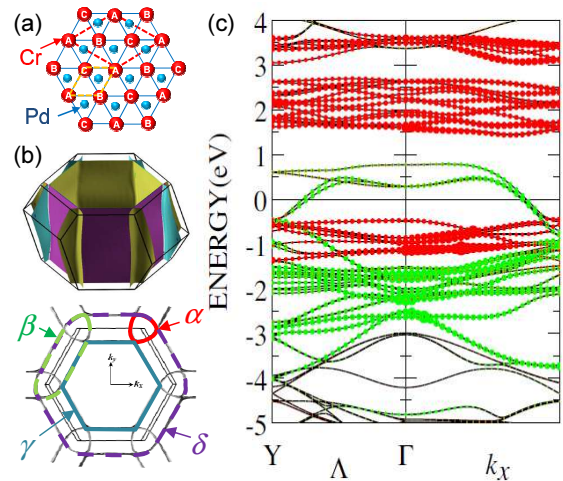}
\caption{\label{fig2}(color online) (a) A triangular Pd layer stacked with a triangular lattice of Cr with three different types of non-coplanar spin directions. The crystal (magnetic) unit cell in indicated by the orange (red) dotted lines. (b) Calculated Fermi surfaces of PdCrO$_2$ and a cross section of the Fermi surface. Edges of the first Brillouin zone are drawn by thin black lines. The triangular ($\alpha$), lens ($\beta$), small hexagonal ($\gamma$) and large hexagonal ($\delta$) orbits are highlighted. The magnetic breakdown orbits ($\beta$ and $\delta$) are indicated by the dashed line.  Band dispersion along the symmetry line $\Lambda$ ($Y$-$\Gamma$) and along the $k_x$ axis. The size of dots is proportional to the character of Pd (green) and Cr (red) states within the muffin-tin
spheres.}
\end{figure}

These FFT results are in strong contrast to those of iso-structural and non-magnetic compound PdCoO$_2$. Recent dHvA experiments on PdCoO$_2$ \cite{PdCoO2:mackenzie:dHvA} reveal two high frequencies at $F$ $\sim$ 30 kT, which related to the local minimum (neck) and the local maximum (belly) in the cross-section of the warped 2D cylindrical FS via the Onsager relation $F$ = ($\hbar c/2\pi e$)$A_k$. The large 2D FS in PdCoO$_2$ is in good agreement with the results of the angle-resolved photoemission spectroscopy and the first principle calculations \cite{PdCoO2:hjnoh:ARPES,PdCoO2:min:band}. For PdCrO$_2$, the largest frequency $\delta$ at $F$ $\sim$ 27.5 kT corresponds to the similar-sized FS as found in PdCoO$_2$. However, several additional smaller frequencies, $\alpha$, $\beta$, and $\gamma$ found in PdCrO$_2$ suggest FS \emph{reconstruction} due to AFM ordering in the CrO$_2$ layer.

In order to identify the origin of the orbits, we performed first principle calculations as shown in Fig. 2. We found negligible difference in total energy between the FM and the AFM interlayer coupling, thus we assumed the ferromagnetic interlayer coupling. The occupied Cr 3$d$ states are of spin-up $t_{2g}$ character as expected. Here the spin axis (up or down) is defined locally at each Cr site so as to be parallel to the atomic-like moment. The spin-up $e_g$ and spin-down $t_{2g}$ states appear at the energies between 1.5 and 2.6 eV, while the spin-down $e_g$ states locate at higher energies, 3-3.7 eV. At the Fermi level $E_F$, two and four bands are crossing the Fermi level along $k_y$ and $k_x$ directions, respectively. Although these conducting states have strong Pd 4$d$ character, Cr 3$d$ states are mixed to some extent. These bands show spin splitting, a profound feature known to be present in spin-spiral antiferromagnets. Although the net magnetization is zero, the exchange fields at Cr sites do induce the spin polarization of conduction electrons whose direction is aligned perpendicular to the spiral plane.

Fermi surface obeys strong two dimensionality as shown in Fig. 2(b). One can notice that there are two types of FS; small triangular pillar with electron character at the corners of the BZ and one big hexagonal pillar with hole character centered at the $\Gamma$ points. Compared to the nonmagnetic PdCoO$_2$, the unit cell becomes 3 times larger due to the ordered Cr spins in a noncollinear 120$^{\rm o}$ spin structure (a rotated $\sqrt{3}\times\sqrt{3}$ supercell) [Fig. 2(a)]. Thus, foldings of the electronic bands into the reduced Brillouin zone (BZ) leads to significant reconstruction of the electronic structure in PdCrO$_2$. The size of reconstructed FS's in fact matches well with the $\alpha$ and $\gamma$ orbits.

Two dimensionality of the FS's is confirmed by the angle dependence of the frequencies ($\alpha$ and $\gamma$) with the tilted magnetic field. As shown in Fig. 3(a) the frequencies for both orbits varies $F$ $\propto$ 1/cos($\theta$), indicating the almost 2D FS with minimal $k_z$-axis dispersion, consistent with the calculations. The number of the orbits detected in experiments is less than that predicted in the calculations. Nevertheless, our results clearly demonstrate that FS reconstruction in PdCrO$_2$ is induced by the 120$^{\rm o}$ ordering of Cr spins.

\begin{figure}
\includegraphics*[width=8.5cm, bb=37 510 365 750]{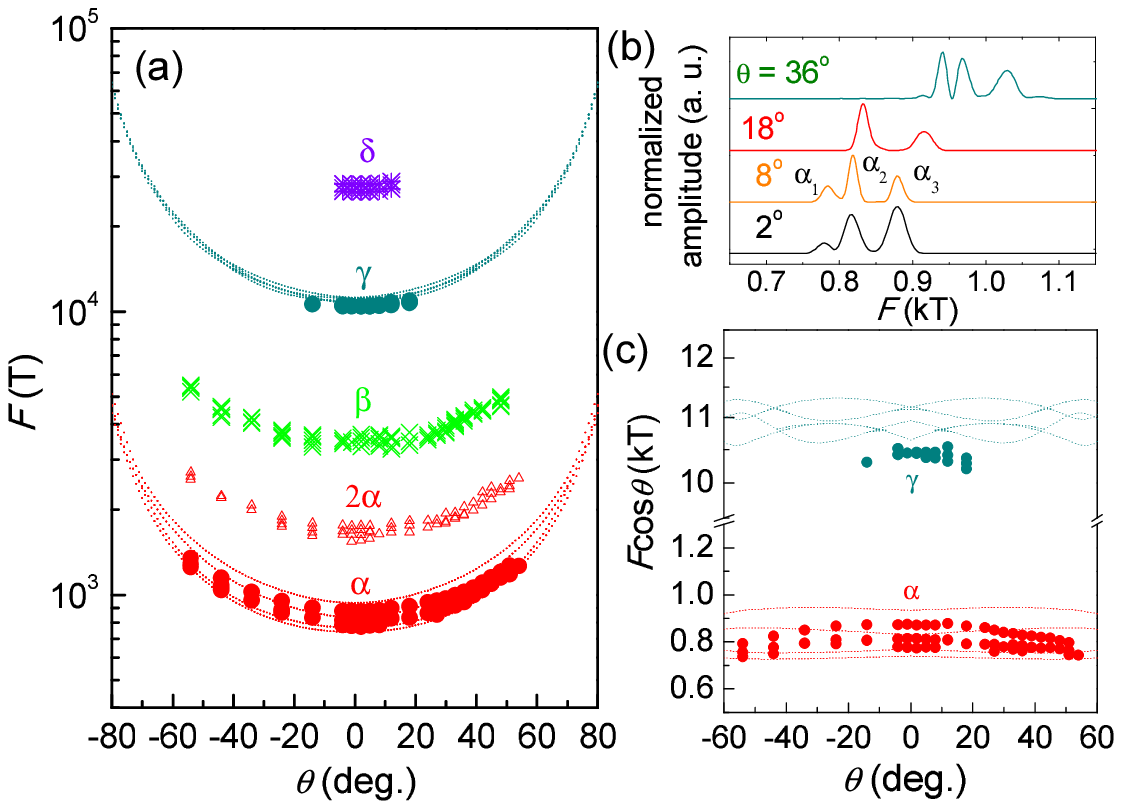}
\caption{\label{fig3}(color online)(c) Angular dependence of all observed dHvA frequencies ($F$) with magnetic field angle $\theta$. $\theta$ is a tilt angle from [001] to [100] direction. The dashed lines show the results of the DFT calculations. (d) The angle dependence of the FFT spectra for the $\alpha$ orbits. (e) The angular dependence of $F\cos\theta$ for the $\alpha$ and the $\gamma$ orbits. }
\end{figure}

The additional frequencies, $\beta$ and $\delta$ branches, are consistent with the sizes of the lenz-type ($\beta$) and the large hexagonal ($\delta$) orbits in Fig. 3(b). In these orbits, the electrons need to tunnel through the gap ($\delta k_g$) in $k$-space from one part of the Fermi surface to another with sufficient cyclotron energy at high magnetic fields, as known as the magnetic breakdown (MB) effects \cite{QO:shoenberg}. For the large hexagonal orbit ($\delta$), the breakdown field $B_0$ is estimated to be $\sim$ 7 T~\cite{supp}. This corresponds to $\delta k_g$ $\sim$ 3.5(2)$\times$ 10$^{-3}$ $\rm{\AA}^{-1}$ using the Chamber formula $B_0$ = $(\pi\hbar/e)(k^3_g/(a+b))^{1/2}$ where $a$ and $b$ are related to the curvatures of the neighboring FS's~\cite{supp}. This is consistent with  $\delta k_g$ $\approx$ 9.7 $\times$ 10$^{-3}$ $\rm{\AA}^{-1}$ from calculations.

Having established that the conduction electrons are strongly coupled to the Cr spins, we now discuss the temperature dependence of dHvA oscillations for the $\alpha$ peak. Since the $\alpha$ FS is sensitive to the band folding induced by the magnetic structure of Cr spins, one can expect that the corresponding dHvA oscillations will be changed if the spin structure is modified as proposed in Ref.~[\onlinecite{PdCrO2:maeno:AHE}]. It has been suggested that in order to induce the finite scalar spin chirality and the corresponding the AHE, the magnetic configuration of the Cr spins should break the $\sqrt{3} \times \sqrt{3}$ periodicity. As shown in Fig. 4(a) and (b), however, we have not found any signature of the changes in the frequencies of the dHvA oscillations as the temperature passes through $T^*$ $\sim$ 20 K.  Also, their amplitude is gradually reduced without any anomaly. This strongly suggests that a change in Cr spin structure at $T^*$ ($<$ $T_N$) is, if any, minute.

\begin{figure}
\includegraphics*[width=8.0cm, bb=0 220 600 750]{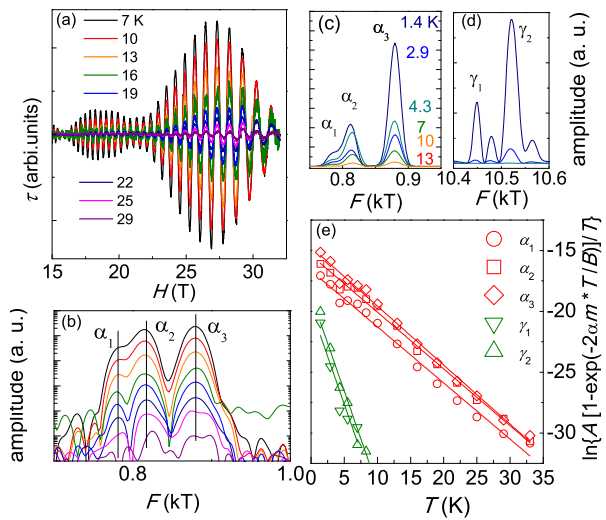}
\caption{\label{fig3}(color online) (a) The oscillatory part of the torque signal as a function of applied magnetic fields at $\theta$ = 5 $^{\rm o}$ for different temperatures. (b) The corresponding FFT spectra of the dHvA oscillations for the $\alpha$ FS in a logarithmic scale. Note that the $\alpha$ FS is the reconstructed FS by the 120$^{\rm o}$ Cr ordering, and thus sensitive to the magnetic structure of Cr spins. The temperature dependence for (c) the $\alpha$ and (d) the $\gamma$ orbits is shown in a linear scale. (e) The mass plot for the $\alpha$ and $\gamma$ orbits.}
\end{figure}

\begin{table}
  \caption{Calculated and measured dHvA frequencies. The measured frequencies are taken from the data at $\theta$ = 5$^{\rm o}$. The calculated ($m_b$) and measured ($m^*$) cyclotron masses are listed in units of the free electron mass ($m_e$).}
  \label{tbl:samples}
  \begin{tabular}{cccccccc}
    \hline
     band  &calc. & & & Exp. &  &  &  \\
     Orbit & $F$ & $m_b$ & & Orbit & $F$ & $m^*$ & $m^*/m_b$$-$1  \\
           & (T) & ($m_e$) & & & (T) & ($m_e$) &   \\
    \hline
     1a & 739 & 0.25 & &  &  &  &  \\
     1b & 769 & 0.26 & & $\alpha_1$ & 783(2) & 0.44(1) &  0.7(1)  \\
     1c & 830 & 0.28 & & $\alpha_2$ & 815(2) & 0.45(1) &  0.4(1)  \\
     1d & 933 & 0.31 & & $\alpha_3$ & 880(1) & 0.47(1) &  0.47(3) \\
     2a & 10648 & 1.26 & & $\gamma_1$ & 10445(3) & 1.5(3) & 0.2(1) \\
     2b & 10954 & 1.22 & & $\gamma_2$ & 10518(3) & 1.5(1) & 0.27(8) \\
     2c & 11127 & 1.13 & & &  &  &    \\
     2d & 11180 & 1.14 & & &  &  &  \\
    \hline
  \end{tabular}
\end{table}

From the temperature dependence of the FFT amplitude as shown in Fig. 4(c) and 4(d), we estimate the strength of the coupling by comparing the measured effective mass ($m^*$) of the quasiparticles with the band structure value ($m_b$). The $m^*$ for the orbits $\alpha$ and $\gamma$ can be extracted using the so-called mass plot based on the Lifshitz-Kosevich formula [Fig. 4(e)]. The resulting mass enhancement factor $\lambda$ = $m^*/m_b$ $-$ 1 is $\sim$ 0.4-0.7 for the $\alpha$ FS's while it is somewhat reduced to be $\sim$ 0.2-0.3 for the $\gamma$ FS's, as listed in Table 1. While the coupling strength is moderate, consistent with the small Sommerfeld coefficient from the specific heat measurements\cite{PdCrO2:maeno:NPD}, the sizable $\lambda$ for PdCrO$_2$ is distinct from the case of PdCoO$_2$ where $m^*$ is similar or even smaller than the calculated $m_b$\cite{PdCoO2:mackenzie:dHvA}. The orbit dependence of $\lambda$ can be understood in terms of the different degree of Cr character in the FS's. In fact, the calculated Cr character on the triangular FS's ($\alpha$) is 60$\%$ larger than that on the hexagonal ($\gamma$) FS~\cite{Cr}. These results, combining with the significant reduction of the resistivity at $T_N$\cite{PdCrO2:maeno:res}, indicate that itinerant electrons are scattered by fluctuations of the Cr spins, which is expected to be strong above $T_N$, but freezed out below $T_N$.

Based on our results, we can conclude that the itinerant electrons are significantly coupled to the 120$^{\rm o}$ helical order below $T_N$. Additional transition accompanied by modification of the Cr spin structure \cite{PdCrO2:maeno:AHE} below $T_N$ is unlikely, and there is little, if any, modification from the ideal 120$^{\rm o}$ spin structure. This implies that spin chirality mechanism is not sufficient to explain the observed unconventional AHE in PdCrO$_2$. Recently, Tomizawa and Kontani suggested that significant AHE can be induced by the orbital Aharonov-Bohm effects even with negligible spin scalar chilality in a non-collinear spin structure\cite{AHE:kontani:AB}. In this model, the conduction electron acquires a finite Berry phase due to the complex $d$-orbital wave function in the presence of non-collinear spin structure. The minimal distortion from a ideal 120$^{\rm o}$ non-collinear spin structure in PdCrO$_2$ cannot induce the sizable AHE through the scalar spin chirality, whereas it might produce dominant AHE via the orbital Aharonov-Bohm effects as proposed in metallic pyrochlore compound, Pr$_2$Ir$_2$O$_7$\cite{AHE:kontani:AB}.

Alternatively, in a non-collinear antiferromagnetic metal, a $k$-dependent spin-splitting of the Fermi surface occur due to broken inversion symmetry by non-collinear spiral magnetic order as shown in Fig. 2. Competition between spin-split hole and electron orbits and the magnetic breakdown between them might be important to explain the complex magnetic field dependence of Hall resistivity in PdCrO$_2$~\cite{PdCrO2:shishido:multiorbits}. Further studies on the Cr spin structures using single crystals as well as the magnetotransport properties with higher magnetic fields are desirable to test these possibilities.

In summary, combining results of de Haas-van Alphen (dHvA) oscillations and first principle calculations, we show that highly metallic electrons mostly in the Pd layers are strongly coupled to the AFM ordering of the Cr layers in triangular antiferromagnet PdCrO$_2$. The non-collinear 120$^{\rm o}$ AFM ordering in the localized Cr moments induces the significant FS reconstruction as compared to the FS of iso-structural nonmagnetic PdCoO$_2$. Considering that the ideal 120$^{\rm o}$ helical ordering has a zero scalar spin chirality, the unconventional AHE found in PdCrO$_2$ cannot be explained in terms of the spin chiral mechanism. This strongly suggests that in order to understand the complex behavior of Hall resistivity of PdCrO$_2$, alternative mechanisms taking into account of the orbital degree of freedom or the multi-FS's are required. Nevertheless, PdCrO$_2$ is an rare example of metallic TAFM's whose unconventional AHE is not fully understood in terms of the spin chirality mechanism.

\begin{acknowledgments}
The authors thank W. Kang, J. H. Shim and M. H. Lee for fruitful discussion. This work was also supported by the BSR (No. 2010-0005669, No. 2012-013838), SRC (2011-0031558), the Max Planck POSTECH/KOREA Research Initiative Program (No. 2011-0031558) by the National Research Foundation of Korea (NRF) funded by the Ministry of Education, Science and Technology. B.I.M. and K.K. acknowledge the support from the NRF project (No. 2009-0079947, 2011-0025237). The work at Japan was supported by the ��Topological Quantum Phenomena�� (No. 22103002) KAKENHI on Innovative Areas from MEXT of Japan.

\end{acknowledgments}


\begin{thebibliography}{00}
\bibitem{Cr} The Cr component on the triangular FS's are estimated inside the muffin-tin sphere of Cr of 1.98 au. in the FM configuration. They are 0.321(0.169) near triangular FS and 0.219(0.083) near hexagonal FS for the up(down) spin.
\end{thebibliography}
\end{document}